\documentclass[preprint,12pt]{elsarticle}

\usepackage{listings}
\usepackage{xcolor}
\usepackage{color}

\usepackage{amssymb}
\usepackage{amsmath}

\usepackage{listings}

\usepackage{tikz}
\usetikzlibrary{shapes.geometric, arrows}

\tikzstyle{startstop} = [rectangle, rounded corners, minimum width=3cm, minimum height=1cm,text centered, draw=black, fill=red!30]

\tikzstyle{io} = [trapezium, 
trapezium stretches=true, 
trapezium left angle=70, 
trapezium right angle=110, 
minimum width=1cm, 
minimum height=1cm, 
text centered, 
  text width=3cm,
draw=black, fill=blue!30]

\tikzstyle{decision} = [diamond, minimum width=1cm, minimum height=1cm, 
text centered, text width=3cm,
draw=black, fill=green!30]

\tikzstyle{arrow} = [thick,->,>=stealth]

\tikzstyle{process} = [rectangle,
 minimum width=3cm, 
 minimum height=1cm, 
 text centered,
  text width=3cm, 
  draw=black, fill=orange!30]

\newcounter{bla}

\journal{Computer Physics Communications}

\def\ep  {\varepsilon}

\begin{document}

\begin{frontmatter}



\title{\texttt{PrecisionLauricella}: package for numerical computation of Lauricella functions depending on a parameter}


\author[a]{M.A. Bezuglov \corref{author}}
\author[a]{B.A. Kniehl}
\author[b,c]{A.I. Onishchenko}
\author[d]{O.L. Veretin}

\cortext[author] {Corresponding author.\\\textit{E-mail address:} bezuglov.ma@phystech.edu}
\address[a]{II.~Institut f\"ur Theoretische Physik, Universit\"at Hamburg, Hamburg, Germany}
\address[b]{Bogoliubov Laboratory of Theoretical Physics, Joint
Institute for Nuclear Research, Dubna, Russia,}
\address[c]{Budker Institute of Nuclear Physics, Novosibirsk, Russia}
\address[d]{Institut f\"ur Theoretische Physik, Universit\"at Regensburg, Regensburg, Germany}

\begin{abstract}
We introduce the \texttt{PrecisionLauricella} package, a computational tool developed in Wolfram Mathematica for high-precision numerical evaluations of Lauricella functions with indices linearly dependent on a parameter, $\ep$. The package leverages a method based on analytical continuation via Frobenius generalized power series, providing an efficient and accurate alternative to conventional approaches relying on multi-dimensional series expansions or Mellin--Barnes representations. This one-dimensional approach is particularly advantageous for high-precision calculations and facilitates further optimization through $\ep$-dependent reconstruction from evaluations at specific numerical values, enabling efficient parallelization. The underlying mathematical framework for this method has been detailed in our previous work, while the current paper focuses on the design, implementation, and practical applications of the \texttt{PrecisionLauricella} package.
\\


\noindent \textbf{PROGRAM SUMMARY}

\begin{small}
\noindent
{\em Program Title:} PrecisionLauricella                                         \\
{\em Developer's repository link:} https://bitbucket.org/BezuglovMaxim/precisionlauricella-package/src/main/ \\
{\em Licensing provisions(please choose one):} GPLv3  \\
{\em Programming language:} Wolfram Mathematica                                    \\
{\em Supplementary material:}  PrecisionLauricella\_Examples.nb                               \\
{\em Nature of problem(approx. 50-250 words):} Lauricella functions, generalizations of hypergeometric functions, appearing in physics and mathematics, including Feynman integrals and string theory. When their indices depend linearly on a parameter $\ep$, their numerical evaluation becomes challenging due to the complexity of high-dimensional series and singularities. Traditional methods, like hypergeometric re-expansion or Mellin--Barnes integrals, often lack efficiency and precision.

Managing multi-dimensional sums exacerbates computational costs, especially for high-precision requirements, making these approaches unsuitable for many practical applications. Thus, there is a pressing need for efficient, scalable methods capable of maintaining numerical accuracy and effectively handling parameter dependencies.\\
{\em Solution method(approx. 50-250 words):} Our method uses the Frobenius approach to achieve analytical continuations of Lauricella functions through generalized power series. Representing the functions as one-dimensional series simplifies high-precision numerical evaluation compared to traditional methods relying on multi-dimensional expansions or Mellin--Barnes integrals.

We further optimize calculations by reconstructing $\ep$ dependencies from evaluations at specific values, enabling efficient parallelization and reducing computational costs.

A comprehensive mathematical exposition of the method is provided in our previous work \cite{1}. \\

\end{small}
   \end{abstract}
\end{frontmatter}






\section{Introduction}
\label{sec:Introduction}

Hypergeometric functions, both in one and several variables, play a fundamental role in physics and mathematics, particularly in quantum field theory and the computation of Feynman integrals. Numerous approaches have been devised to express Feynman integrals through hypergeometric functions, including the Mellin--Barnes method,\footnote{For an introduction and references to foundational works, see Refs.~\cite{Weinzierl:2022eaz,Dubovyk:2022obc,smirnov2006feynman}.} the DRA method \cite{Tarasov:2006nk, Lee:2009dh, Lee:2012hp}, based on dimensional recurrence relation and analytical properties, the method of functional equations \cite{Tarasov:2022clb}, the exact Frobenius method \cite{Bezuglov:2022npo, Bezuglov:2021tax, Blumlein:2021hbq}, and the Gelfand--Kapranov--Zelevinsky (GKZ) approach to Feynman integrals\footnote{For a comprehensive overview, see Refs.~\cite{Matsubara-Heo:2023ylc,Vanhove:2018mto}.} \cite{GKZ1,GKZ2,GKZ3,GKZ4,GKZ5,beukers2013monodromy,Kalmykov:2012rr,delaCruz:2019skx,Klausen:2019hrg,Ananthanarayan:2022ntm}. These hypergeometric functions often depend on parameters linearly related to the dimensional regularization parameter $\ep$, and their Laurent expansions in $\ep$ are essential for practical applications. Several computational tools have been developed for these expansions, both numerically and analytically\footnote{See also \cite{Greynat:2014jsa,Greynat:2013hox,Yost:2011wk,Moch:2001zr,Davydychev:2003mv,Kalmykov:2006pu,Kalmykov:2006hu,Kalmykov:2007pf,Kalmykov:2010gb} and references therein for additional works on $\ep$-expansion of hypergeometric functions.} \cite{Huber:2005yg,Huber:2007dx,Moch:2005uc,Weinzierl:2002hv,Ablinger:2013cf,Huang:2012qz,Bera:2023pyz,Bezuglov:2023owj}.

This paper introduces \texttt{PrecisionLauricella}, a software package written in Wolfram Mathematica, designed for the high-precision numerical evaluation of Lauricella functions with indices linearly dependent on the parameter $\ep$. The mathematical foundations of the algorithm implemented in this package have been detailed in our previous work \cite{npb}. The package enables computations of these functions' Laurent expansions about $\ep = 0$ for arbitrary argument values. The primary challenge lies in the analytical continuation of the defining series representations of Lauricella functions from their convergence domains to the entire $\mathbb{C}^n$ space, where $n$ is the number of arguments. While classical integral representations such as those of Euler \cite{Exton} and Mellin--Barnes \cite{Exton,AppelKampedeFeriet} can, in principle, address this continuation, they are less practical for high-precision numerical evaluations. Instead, series representations that solve the corresponding partial differential equations and are valid in subdomains covering the entire $\mathbb{C}^n$ space are more effective. These representations not only facilitate accurate numerical computations, but also align with the classical approaches to analytic continuation developed by Kummer and Riemann \cite{AnalyticCont1,AnalyticCont2}, which emphasize the monodromy group.

Existing methods for the analytic continuation of Lauricella functions can be categorized into two main approaches: re-expansion of hypergeometric series using known continuations of simpler functions \cite{AppelKampedeFeriet,Erdelyi,Olsson,Exton,Ananthanarayan:2024nsc,Bera:2024hlq} and Mellin--Barnes integral representations \cite{hahne1969analytic,Bezrodnykh1,Bezrodnykh2,Bezrodnykh3,Bezrodnykh4,Bezrodnykh5,BezrodnykhReview,HYPERDIRE1,HYPERDIRE2,HYPERDIRE3,HYPERDIRE4}. 

In this work, we address the analytic continuation problem by leveraging the Frobenius method to construct generalized power series solutions of the Pfaffian systems satisfied by Lauricella functions along chosen analytic-continuation paths. This approach produces one-dimensional series, which are computationally more efficient than the multi-dimensional series arising in other methods, such as Mellin--Barnes representations. \texttt{PrecisionLauricella} automates the entire process, offering an accessible and efficient tool for researchers dealing with Lauricella functions and their applications.

The remainder of the paper is organized as follows. In Section~\ref{sec:LauricellaFunctions}, we briefly recall the definition of Lauricella functions and associated systems of differential equations.
Section~\ref{sec:Algorithm}, we provide a brief description of all the steps of the algorithm and present the program's flowchart.
Section~\ref{sec::PrecisionLauricella} contains details on the usage of the \texttt{PrecisionLauricella} package and its performance.
Finally, in Section~\ref{sec:Conclusion}, we present our conclusion.

\section{Lauricella Functions and associated differential equations}
\label{sec:LauricellaFunctions}
In this section, we discuss the class of functions that are implemented in the presented package. Specifically, we focus on three Lauricella functions: $F_A^{(n)}$, $F_B^{(n)}$, and $F_D^{(n)}$ for $n \leq 3$. Generally, these functions are defined as hypergeometric series of several variables \cite{bateman1953higher,Schlosser:2013hbz}:
\begin{eqnarray}
\lefteqn{F_A^{(n)}(\alpha; \beta_1,\dots, \beta_n; \gamma_1,\dots,\gamma_n; x_1,\dots,x_n)}
\nonumber\\
&=&\sum\limits_{m_1,\dots, m_n = 0}^{\infty} \frac{(\alpha)_{m_1+\dots+m_n} (\beta_1)_{m_1} \dots (\beta_n)_{m_n}}{(\gamma_1)_{m_1} \dots (\gamma_n)_{m_n} \, m_1! \dots m_n!} \, x_1^{m_1} \dots x_n^{m_n}\,,
\nonumber\\
\lefteqn{F_B^{(n)}(\alpha_1,\dots,\alpha_n; \beta_1,\dots, \beta_n; \gamma; x_1,\dots,x_n)} 
\nonumber\\
&=&\sum\limits_{m_1,\dots, m_n = 0}^{\infty} \frac{(\alpha_1)_{m_1} \dots (\alpha_n)_{m_n} (\beta_1)_{m_1} \dots (\beta_n)_{m_n}}{(\gamma)_{m_1+\dots+m_n} \, m_1! \dots m_n!} \, x_1^{m_1} \dots x_n^{m_n}\,,
\nonumber\\
\lefteqn{F_D^{(n)}(\alpha; \beta_1,\dots, \beta_n; \gamma; x_1,\dots,x_n)}
\nonumber\\
&=&\sum\limits_{m_1,\dots, m_n = 0}^{\infty} \frac{(\alpha)_{m_1+\dots+m_n} (\beta_1)_{m_1} \dots (\beta_n)_{m_n}}{(\gamma)_{m_1+\dots+m_n} \, m_1! \dots m_n!} \, x_1^{m_1} \dots x_n^{m_n}\,.
\end{eqnarray}
All these series converge absolutely within specific regions. For the functions $F_B^{(n)}$ and $F_D^{(n)}$, the region of convergence is given by $|x_1| < 1$, $|x_2| < 1$, \dots, $|x_n| < 1$. For the function $F_A^{(n)}$, the region is given by $|x_1| + |x_2| + \dots + |x_n| < 1$.

When reduced to a single variable, these series correspond to the well-known hypergeometric function:

\begin{equation}
F_A^{(1)} = F_B^{(1)} = F_D^{(1)} = {}_2F_1\,.
\end{equation}

Calculating these functions within their convergence regions is straightforward and does not pose significant challenges. However, our primary interest lies in their analytic continuation. To achieve this, we utilize partial differential equations associated with these functions. These partial differential equations can be derived directly from the series definitions.

Consider a general series of the form 
\begin{equation}
f = \sum A_{i_1,i_2,\dots,i_n} x_1^{i_1} x_2^{i_2} \dots x_n^{i_n}\,,
\end{equation}
where 
\begin{equation}
\frac{A_{i_1,\dots,i_k+1,\dots,i_n}}{A_{i_1,\dots,i_k,\dots,i_n}} = \frac{P_k(i_1,i_2,\dots,i_n)}{P'_k(i_1,i_2,\dots,i_n)}\,,
\end{equation}
and $P_k$, $P'_k$ are polynomials in their variables. The function $f$ then satisfies the following system of partial differential equations:
\begin{equation}
\left[P'_k\left(\theta_{x_{1}}, \theta_{x_{2}}, \dots, \theta_{x_{n}}\right) \frac{1}{x_k} - P_k\left(\theta_{x_{1}}, \theta_{x_{2}}, \dots, \theta_{x_{n}}\right)\right] f = 0\,, \qquad k = 1, \dots, n\,,
\end{equation}
where $\theta_a = a \partial/\partial a$.

For example, for the function $F_B^{(2)}(\alpha_1, \alpha_2; \beta_1, \beta_2; \gamma; x, y) = F_3$, the partial differential equations are:
\begin{eqnarray}
\small
\left[x(1-x)\frac{\partial^2}{\partial x^2} + y\frac{\partial^2}{\partial x \partial y} + \left(\gamma - (\alpha_1 + \beta_1 + 1)x\right)\frac{\partial}{\partial x} - \alpha_1 \beta_1 y \frac{\partial}{\partial y} - \alpha \beta_1\right] F_3  &=& 0\,,
\nonumber\\
\left[y(1-y)\frac{\partial^2}{\partial y^2} + x\frac{\partial^2}{\partial x \partial y} + \left(\gamma - (\alpha_2 + \beta_2 + 1)y\right)\frac{\partial}{\partial y} - \alpha_2 \beta_2 x \frac{\partial}{\partial x} - \alpha \beta_2\right] F_3  &=& 0\,.
\nonumber\\
&&
\end{eqnarray}
While obtaining such partial differential systems is straightforward, applying the Frobenius method to them can be challenging. To facilitate this, we require a complete system of equations in Pfaffian form:
\begin{equation}
dJ = \left(M_x \, dx + M_y \, dy\right) J\,.
\end{equation}
If we choose a basis as
\begin{equation}
J = \left\{F_3,~ \theta_x F_3,~ \theta_y F_3, ~\theta_x\theta_y F_3\right\}\,, 
\end{equation}
then the matrices will have the form
\begin{eqnarray}
\scriptsize
M_x&=&\left(
\begin{array}{cccc}
 0 & \frac{1}{x} & 0 & 0 \\
 -\frac{\alpha_1 \beta_1}{x-1} & -\frac{-\gamma +\alpha_1
   x+\beta_1 x+1}{(x-1) x} & 0 & \frac{1}{(x-1) x} \\
 0 & 0 & 0 & \frac{1}{x} \\
 0 & -\frac{\alpha_2 \beta_2 y}{x (x y-x-y)} & -\frac{\alpha_1
   \beta_1 (y-1)}{x y-x-y} & -\frac{-\alpha_1 x-\beta_1
   x+\alpha_1 x y+\beta_1 x y+\alpha_2 y+\beta_2 y-\gamma 
   y+y}{x (x y-x-y)} \\
\end{array}
\right)
\,,
\nonumber\\
M_y&=&
\left(
\begin{array}{cccc}
 0 & 0 & \frac{1}{y} & 0 \\
 0 & 0 & 0 & \frac{1}{y} \\
 -\frac{\alpha_2 \beta_2}{y-1} & 0 & -\frac{-\gamma +\alpha_2
   y+\beta_2 y+1}{(y-1) y} & \frac{1}{(y-1) y} \\
 0 & -\frac{\alpha_2 \beta_2 (x-1)}{x y-x-y} & -\frac{\alpha_1
   \beta_1 x}{y (x y-x-y)} & -\frac{\alpha_1 x+\beta_1 x-\gamma 
   x+\alpha_2 x y+\beta_2 x y+x-\alpha_2 y-\beta_2 y}{y (x
   y-x-y)} \\
\end{array}
\right)\,.
\nonumber\\
&&
\normalsize
\end{eqnarray}
In our work, for the functions under consideration, we utilize pre-derived Pfaffian systems from Ref.~\cite{Bezuglov:2023owj} with the following function bases:
\begin{eqnarray}
&&\left\{\theta_{x_{j_1}} \dots \theta_{x_{j_k}} F_N \,\Big|\, 0 \leq k \leq n,\, j_1 < j_2 < \dots < j_k \right\}\,, \qquad N = A, B\,,
\nonumber\\
&&\left\{F_D,\, \theta_{x_{j}} F_D \,\Big|\, j = 1, \dots, n \right\}\,.
\end{eqnarray}

Our aim is to evaluate the value of a chosen Lauricella function at the point $\vec{x}$ given its initial value at the point $\vec{x}_0$. This can be most easily done by considering the above differential equation system along some path connecting these two points. Suppose we want to get a solution along the path $f$  parameterized by the parameter $t$, such that 
\begin{equation}
x_1 = f_1(t)\,,\quad x_2 = f_2(t)\,,\quad \dots\,,\quad x_n = f_n(t)\,.
\end{equation}
In this way, the differential equation system in Pfaffian form restricted to this path takes the form
\begin{equation}
\frac{dJ}{dt} = M_t J\,, \qquad M_t = \frac{\partial x_1}{\partial t}M_1+\frac{\partial x_2}{\partial t}M_2+\dots+\frac{\partial x_n}{\partial t}M_n\,.
\end{equation}
In our work, we choose the contour $f$ as a line passing through the center of coordinates,
\begin{equation}
x_1 = \kappa_1 t\,,\quad x_2 = \kappa_2 t\,,\quad \dots\,,\quad x_n = \kappa_{n} t\,.
\label{eq:kappas}
\end{equation}

\section{Algorithm overview}
\label{sec:Algorithm}
\begin{figure}[h]
\scalebox{0.46}{\begin{tikzpicture}[node distance=2cm]

\draw [draw=black, line width=1pt, dashed] (12.5,-5.0) rectangle (26,-14);
\draw [draw=black, line width=1pt, dashed] (7.5,-11.0) rectangle (12.1,-20);

\node[text centered, text width=8cm] at (20,-15) {This block is responsible for finding the path of analytical continuation of the series beyond the convergence region};

\node[text centered, text width=3cm] at (2,-12) {This loop can run no more than four times};

\node[text centered, text width=4cm] at (14.5,-17.5) {This block can be efficiently parallelized};

\node (start) [startstop] {Start};
\node (in1) [io, right of=start, xshift=3.0cm] {Target function, precision and expansion data};
\node (pro1) [process, right of=in1, xshift=3.0cm] {boundary conditions
and $\ep$-lattice};
\node (pro2) [process, below of=pro1, yshift=-0.5cm] {Derive a system of differential equations in one variable};
\node (in2) [io, right of=pro2, xshift=3.5cm] {System of differential
equations in Pfaffian form};
\node (dec1) [decision, below of=pro2, yshift=-2.5cm] {Need an analytic
continuation?};
\node (anal1) [process, right of=dec1, xshift=3.0cm] {generate a set of regions on a grid};
\node (anal2) [process, right of=anal1, xshift=2.0cm] {Constract intersection
graph};
\node (anal3) [process, right of=anal2, xshift=2.0cm] {Find paths connecting
start and end points in a
graph};
\node (decanal) [decision, below of=anal3, yshift=-2.0cm] {Is there at least
one path?};
\node (anal4) [process, left of=decanal, xshift=-3.0cm] {increase the grid step and
enlarge the generation
area};
\node (pro3) [process, below of=dec1, yshift=-4.5cm] {Derive recurrence relations
for Frobenius series coefficients for all regions along the analytic continuation};
\node (pro4) [process, below of=pro3, yshift=-2.5cm] {Evaluate the U matrix at
each matching point along
the analytical continuation
path};
\node (dec2) [decision, below of=pro4, yshift=-3.0cm] {Have we reached
the end point?};
\node (pro5) [process, left of=dec2, xshift=-4.0cm] {Move the origin of the
system to the current point};
\node (pro6) [process, right of=dec2, xshift=4.0cm] {Perform the reconstruction
of Laurent series expansion
in $\ep$ using Lagrange
interpolation polynomials};

\node (out1) [io, right of=pro6, xshift=3.0cm] {Output};
\node (stop) [startstop, right of=out1, xshift=3.0cm] {Stop};

\draw [arrow] (start) -- (in1);
\draw [arrow] (in1) -- (pro1);
\draw [arrow] (pro1) -- (pro2);
\draw [arrow] (in2) -- (pro2);
\draw [arrow] (pro2) -- (dec1);
\draw [arrow] (dec1) -- node[anchor=south] {yes} (anal1);
\draw [arrow] (dec1) -- node[anchor=west] {no} (pro3);
\draw [arrow] (anal1) -- (anal2);
\draw [arrow] (anal2) -- (anal3);
\draw [arrow] (anal3) -- (decanal);
\draw [arrow] (decanal) -- node[anchor=south] {no} (anal4);
\draw [arrow] (anal4) -| (anal1);
\draw [arrow] (decanal) |- node[anchor=west] {yes} (pro3);
\draw [arrow] (pro3) -- (pro4);
\draw [arrow] (pro4) -- (dec2);
\draw [arrow] (dec2) -- node[anchor=south] {no} (pro5);
\draw [arrow] (pro5) |- (pro2);
\draw [arrow] (dec2) -- node[anchor=south] {yes} (pro6);
\draw [arrow] (pro6) -- (out1);
\draw [arrow] (out1) -- (stop);
\end{tikzpicture}}
\caption{Flowchart for the \texttt{PrecisionLauricella} package.}
\label{fig:flowchart}
\end{figure}
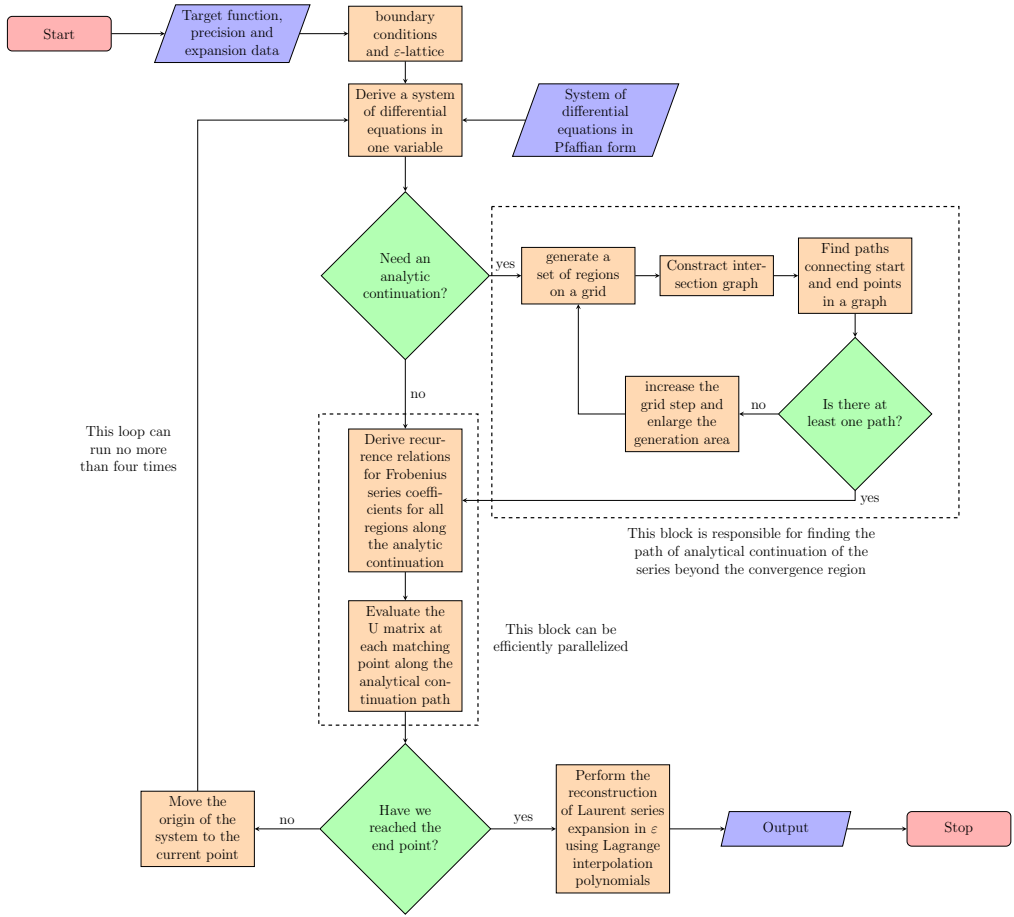

In this section, we provide a general overview of the algorithm and its main stages. The flowchart of the algorithm is presented in Fig.~\ref{fig:flowchart}. The algorithm takes as input the target function, the required precision, and the desired number of terms in the $\ep$ expansion. One of the key features of the algorithm is its treatment of $\ep$ dependencies. Instead of directly expanding the system of equations as a series in $\ep$, we evaluate the system for several numerical values of $\ep$ at specific grid points. The grid is constructed based on the required accuracy and the desired order of expansion in $\ep$. All computations are then performed independently for each grid point, simplifying the calculations and enabling efficient parallelization. In the initial step, we determine the parameters of the grid along with other essential solution parameters, such as boundary conditions.

Next, we select a path in the multivariable space, as described in the previous section. During this step, we impose the condition that all $\kappa_i$ values must be non-negative real numbers. As a result, this step may need to be repeated several times, albeit never more than four times. The rationale behind this condition is discussed in detail in Ref.~\cite{npb}. The system of equations in Pfaffian form is preconfigured in the program for the considered Lauricella functions, which simplifies this process. This step ultimately reduces the problem to a system of differential equations in a single variable.

\begin{figure}[h!]
\centering
\begin{minipage}{0.35\textwidth}
\includegraphics[width=\textwidth]{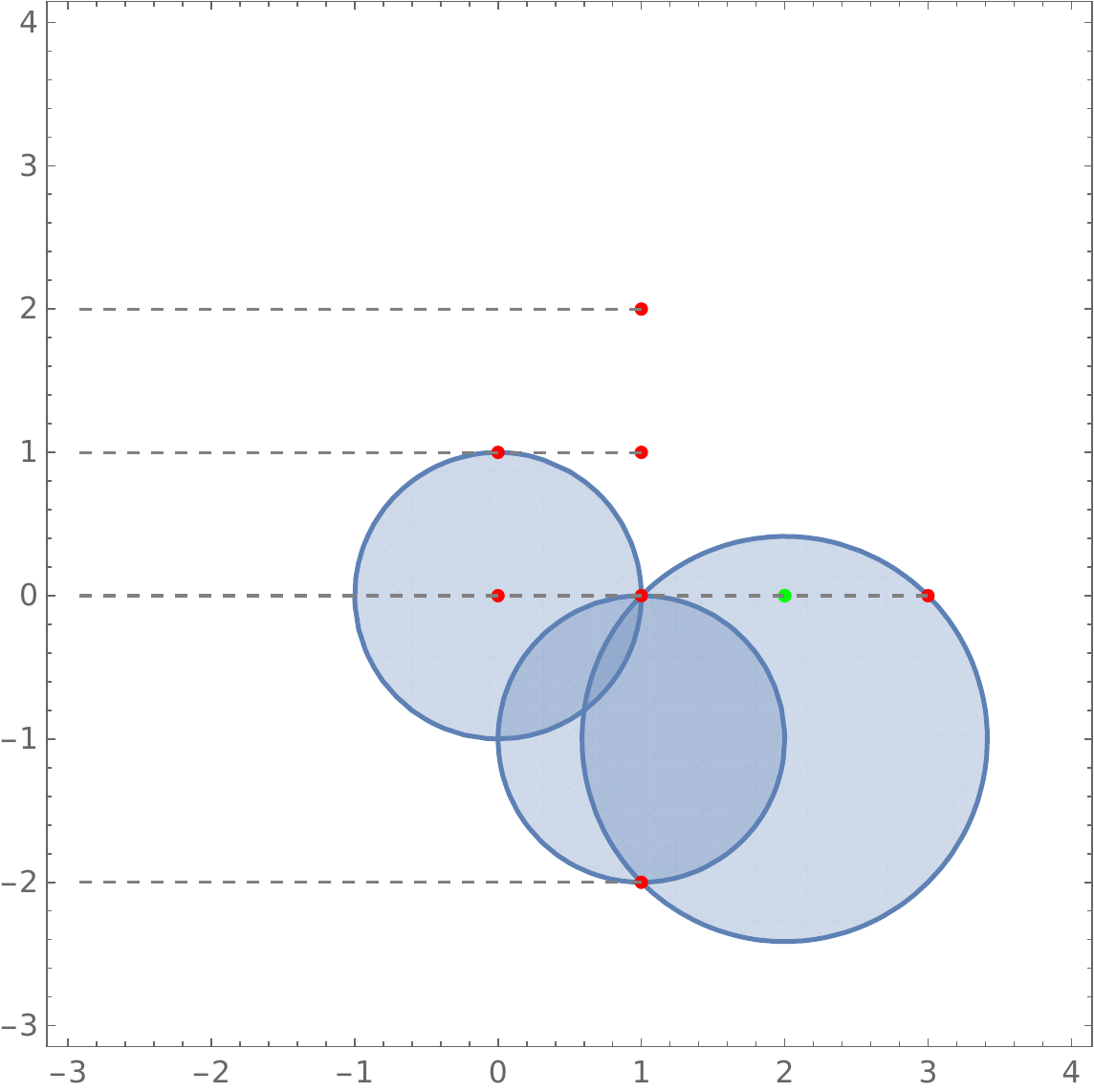}
\label{graph1}
\end{minipage}\hfill
\begin{minipage}{0.55\textwidth}
\includegraphics[width=\textwidth]{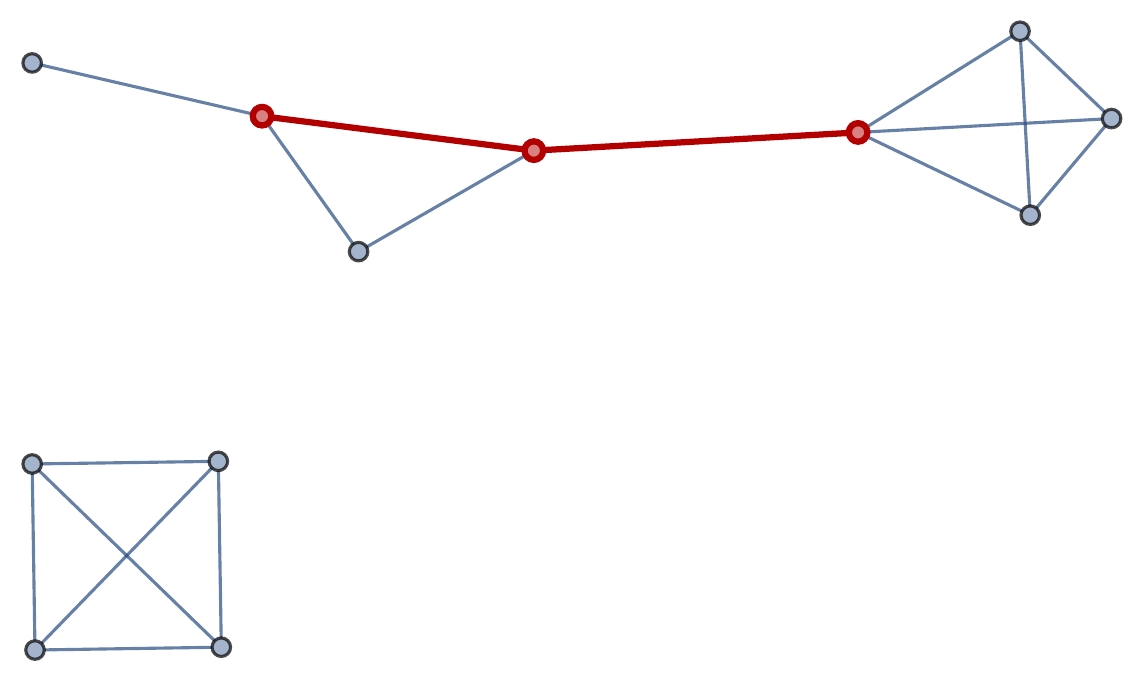}
\label{graph2}
\end{minipage}\vfill
\centering
\begin{minipage}{0.35\textwidth}
\includegraphics[width=\textwidth]{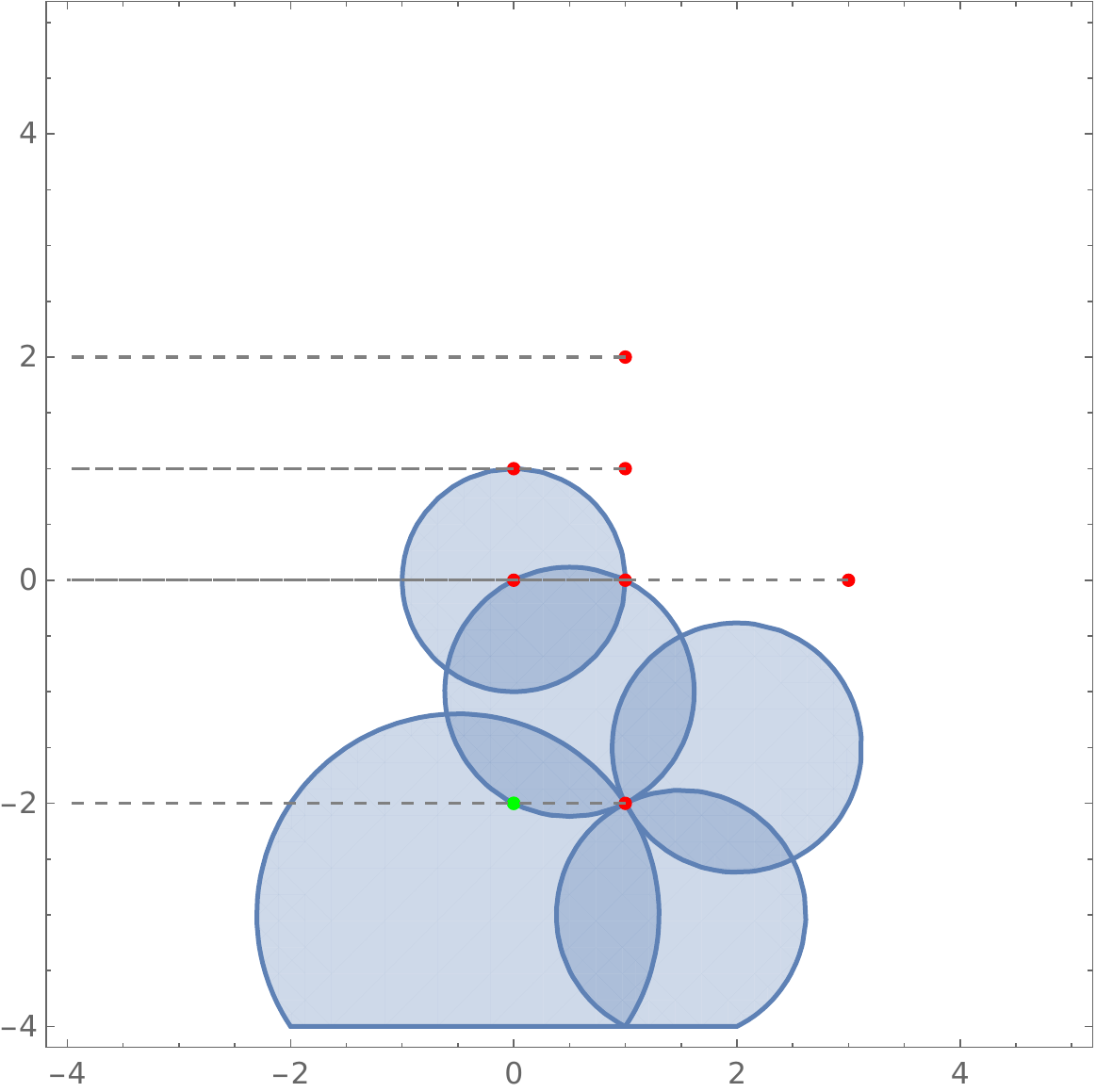}
\label{graph3}
\end{minipage}\hfill
\begin{minipage}{0.55\textwidth}
\includegraphics[width=\textwidth]{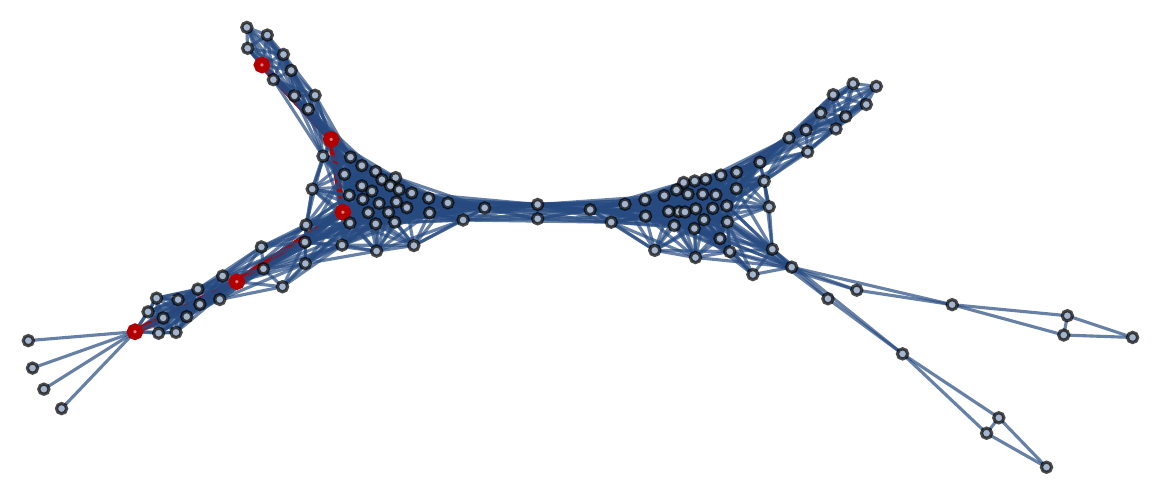}
\label{graph4}
\end{minipage}\vfill
\caption{On the left, one of the possible paths with complex singular points is shown. The singular points are shown in red, the end point in green, the staring point is at the origin, and shaded circles represent convergence regions. Horizontal dotted lines indicate branch cuts of singular points. On the right, the intersection graph of the expansion regions is displayed. The nodes in the intersection graph include expansion regions at regular points. The nodes and edges highlighted in red show the found path of analytic continuation. Note also, that the intersection graph may have several connected components.}
\label{fig:analExample1}
\end{figure}

\begin{figure}[h!]
\centering
\begin{minipage}{0.35\textwidth}
\includegraphics[width=\textwidth]{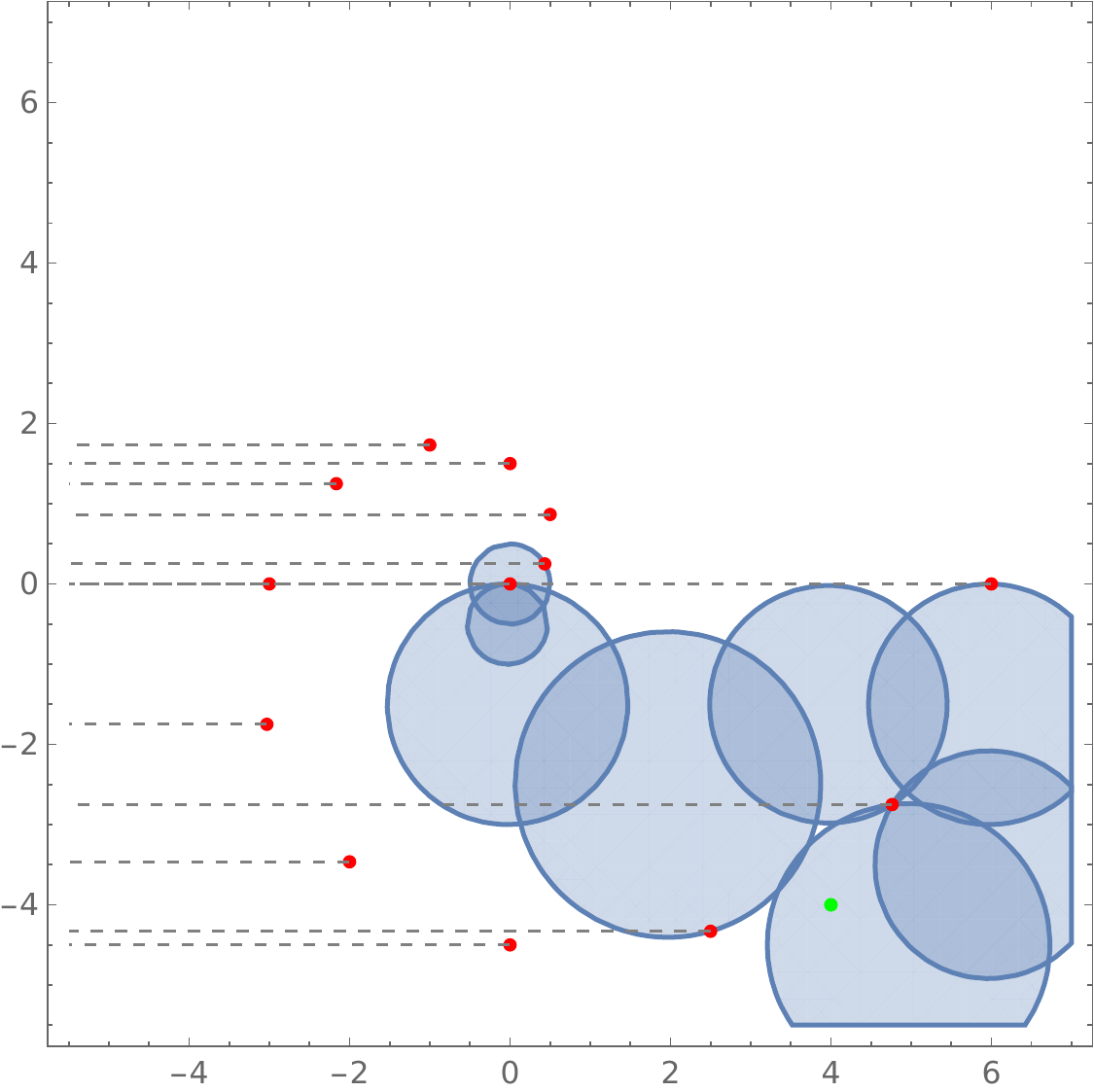}
\label{graph5}
\end{minipage}\hfill
\begin{minipage}{0.55\textwidth}
\includegraphics[width=\textwidth]{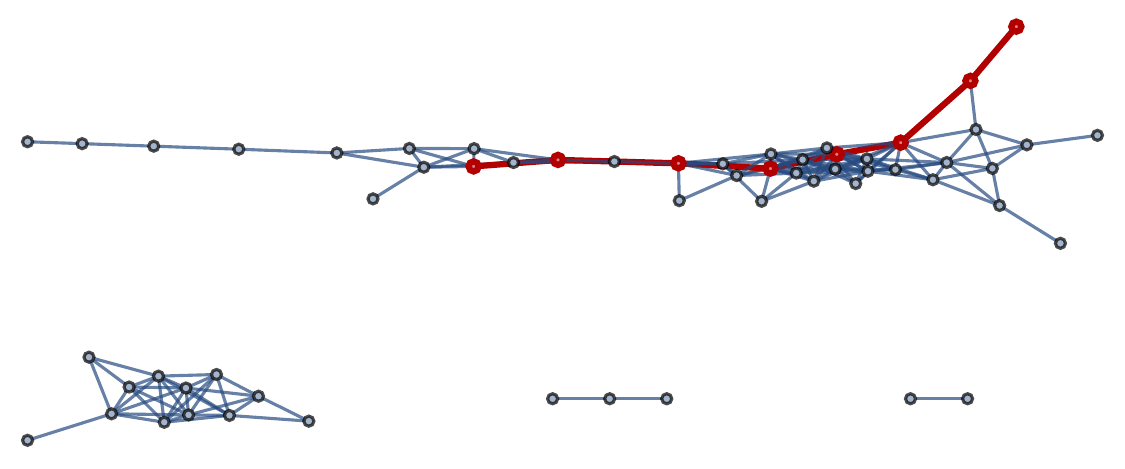}
\label{graph6}
\end{minipage}\vfill
\centering
\begin{minipage}{0.35\textwidth}
\includegraphics[width=\textwidth]{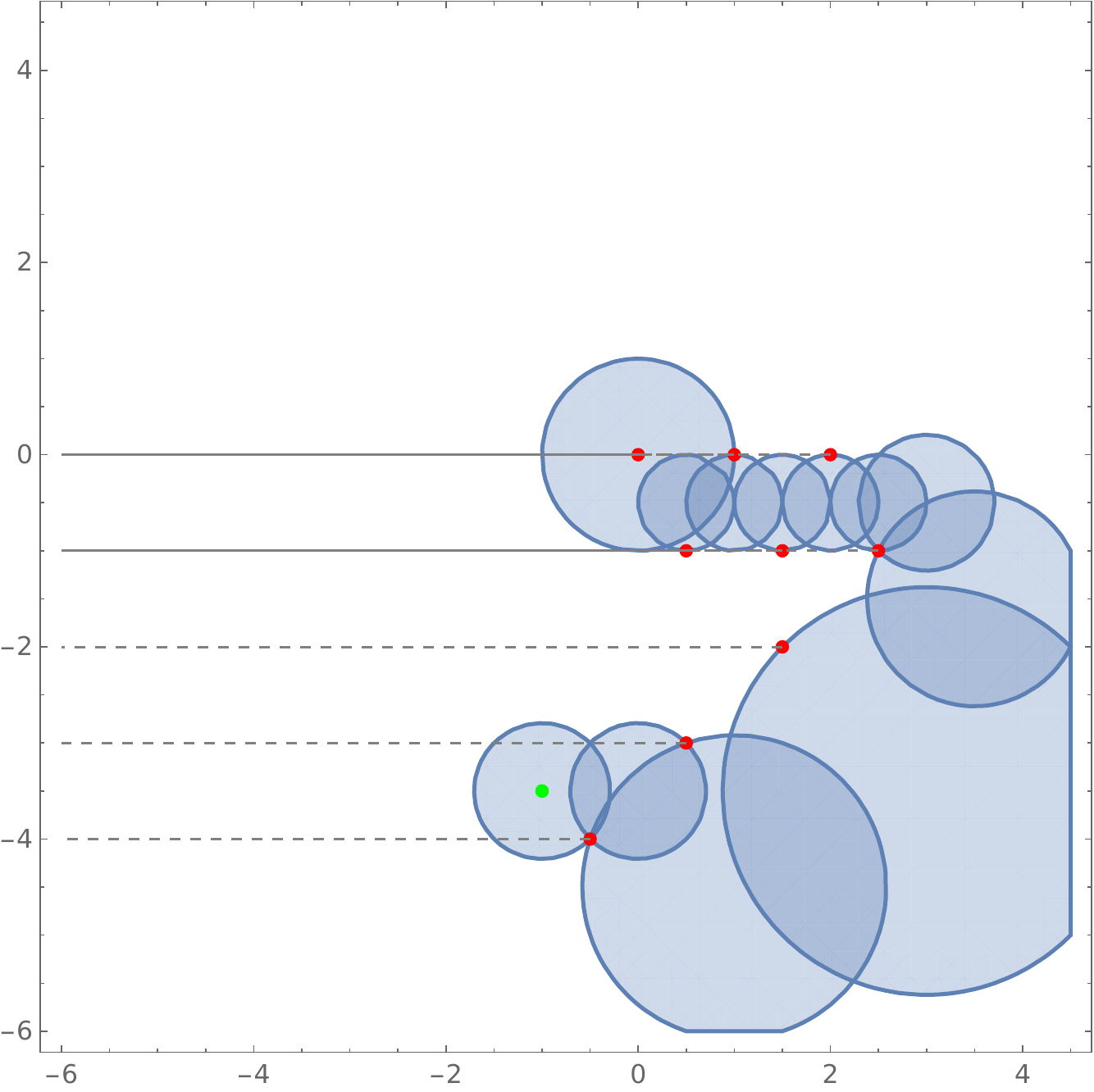}
\label{graph7}
\end{minipage}\hfill
\begin{minipage}{0.55\textwidth}
\includegraphics[width=\textwidth]{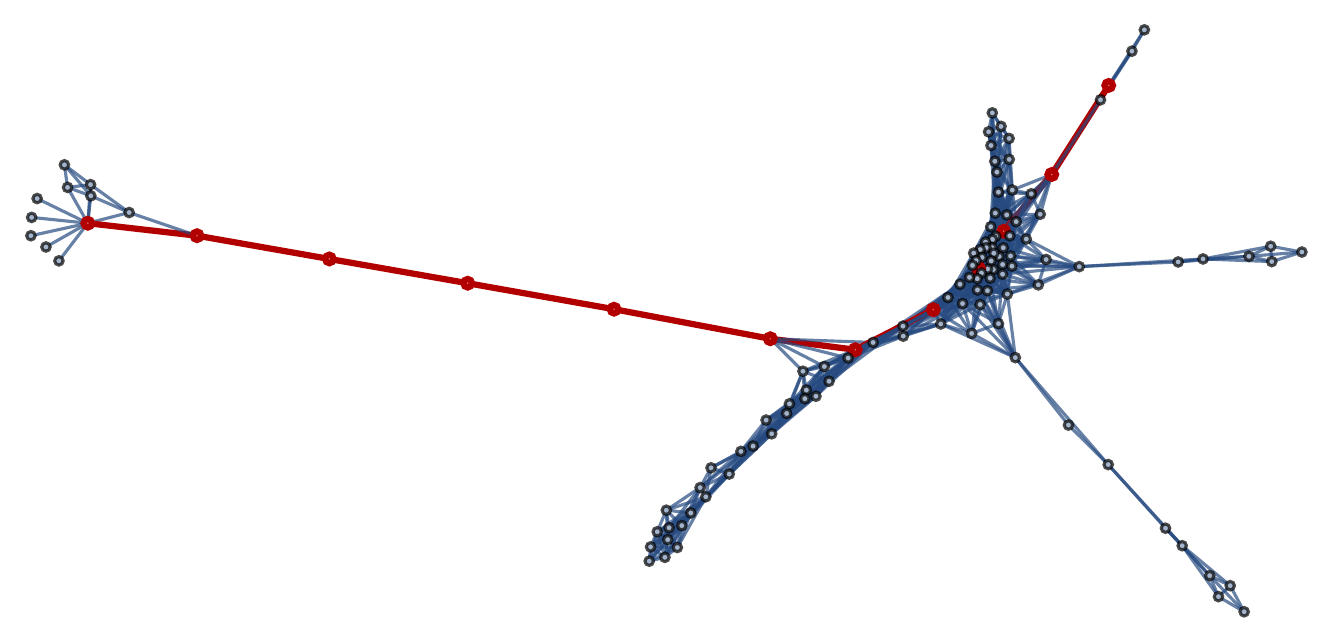}
\label{graph8}
\end{minipage}
\caption{On the left, one of the possible paths with complex singular points is shown. The singular points are shown in red, the end point in green, the staring point is at the origin, and shaded circles represent convergence regions. Horizontal dotted lines indicate branch cuts of singular points. On the right, the intersection graph of the expansion regions is displayed. The nodes in the intersection graph include expansion regions at regular points. The nodes and edges highlighted in red show the found path of analytic continuation. Note also, that the intersection graph may have several connected components.}
\label{fig:analExample2}
\end{figure}

We then determine whether analytic continuation is required. If so, the analytic continuation is performed according to the following procedure. First, we generate circular regions within a rectangular area containing the starting and endpoint. The radii of these regions are determined by their distance from the nearest singularities. Next, we compute the intersection graph: each region is treated as a node, with special nodes representing the starting and end points. Nodes are connected if the minimum distance between regions is less than three quarters of the radius of the region. Branch cuts are also taken into account, with regions on opposite sides of a cut considered non-intersecting. The intersection graph is then used to reduce the continuation problem to a graph traversal between the starting and end nodes, which is solved using a built-in Wolfram Mathematica function. If no path is found, the grid is refined, and the process is repeated until a path is identified. Examples of this procedure are shown in Figs.~\ref{fig:analExample1} and \ref{fig:analExample2}.

Following this, we compute solutions to the systems of differential equations in each region along the analytic continuation path. These systems have the general form
\begin{equation}
\label{eq:genEq}
\frac{d J}{dt}=M(t)J\,.
\end{equation}
Due to our choice of basis for the hypergeometric functions, we can always write the matrix of the differential equation in the form
\begin{equation}
M(t) = \frac{A_0}{t} + R(t)\,,
\end{equation}
where $R(t)$ is a rational function which has no poles at the point $t = 0$.
We are looking for the fundamental solution matrix for each of these equations in the form
\begin{equation}
U = \sum\limits_{\lambda \in S}t^{\lambda}\sum\limits_{n = 0}^{\infty}\sum_{k = 0}^{m_{\lambda}}c_{n}(\lambda, k) t^n \log^k(t)\,,
\label{eq:FrobeniusAn}
\end{equation}
where the set $S$ consists of the non-resonant $A_0$ matrix eigenvalues and $m_{\lambda}$ is the number of eigenvalues in resonance with eigenvalue $\lambda$, that is their differences are integers. If several eigenvalues are in resonance, then the set $S$ will include only the smallest of them. Accordingly, the next two steps consist of deriving difference equations for the coefficients of the Frobenius series and solving these equations. From these solutions, one can obtain a numerical expression for the fundamental solution matrix for each of the regions of analytic continuation and each $\ep$ value from the lattice. 

Finally, we check whether we have reached the end point in the multi-variable plane. If not, we shift the origin of the system to the current point and repeat the previous steps again. If we have reached the end point, we collect all the fundamental solution matrices together to obtain the answer for the end point. Finally, we reconstruct the Laurent series coefficients for the $\ep$ expansion using Lagrange interpolation polynomials.

We can estimate the numerical error of our method as
\begin{equation}
\text{max}\left(h^{2n - 2\lfloor  k/2 \rfloor} , \Delta_{\text{Frob}}\right)\,,
\end{equation}
where $h$ is the $\ep$ lattice step size, $2n$ is the number of lattice elements, $k$ is the number of terms in the $\ep$ expansion and $\Delta_{\text{Frob}}$ is the error arising from the truncation of the Frobenius generalized power series.

\section{PrecisionLauricella package}
\label{sec::PrecisionLauricella}
The \texttt{PrecisionLauricella} package can be freely downloaded from the bitbucket repository https://bitbucket.org/BezuglovMaxim/precisionlauricella-package/src/main/. The entire package consists of one file PrecisionLauricella.wl and, provided the path is set correctly, is loaded with the command 
\begin{lstlisting}[language=Mathematica]
<< PrecisionLauricella`
\end{lstlisting}
The universal function for the numerical expansion of Lauricella functions in Laurent series in parameter $\ep$ is \texttt{NExpandHypergeometry}. This function takes three arguments. The first is a hypergeometric function that needs to be expanded. The second is the list $\{\ep, k\}$, where the first argument $\ep$ is the variable with respect to which the expansion is performed, and the second is the required number of terms in the $\ep$ expansion. The last argument specifies the accuracy of the result with the desired number of decimal places. 
For those functions that are not included by default in Wolfram Mathematica, we introduce our own notations:  \texttt{AppellF2}, \texttt{AppellF3}, \texttt{LauricellaFA}, \texttt{LauricellaFB} and \texttt{LauricellaFD}.  The arguments of these functions are the same as in their definitions. In the case of Lauricella functions, numbered indices and variables are collected into lists.  More details about the package functionality can be found in the Mathematica notebook with examples.

\begin{figure}[h!]
\centering
\begin{minipage}{0.45\textwidth}
\includegraphics[width=\textwidth]{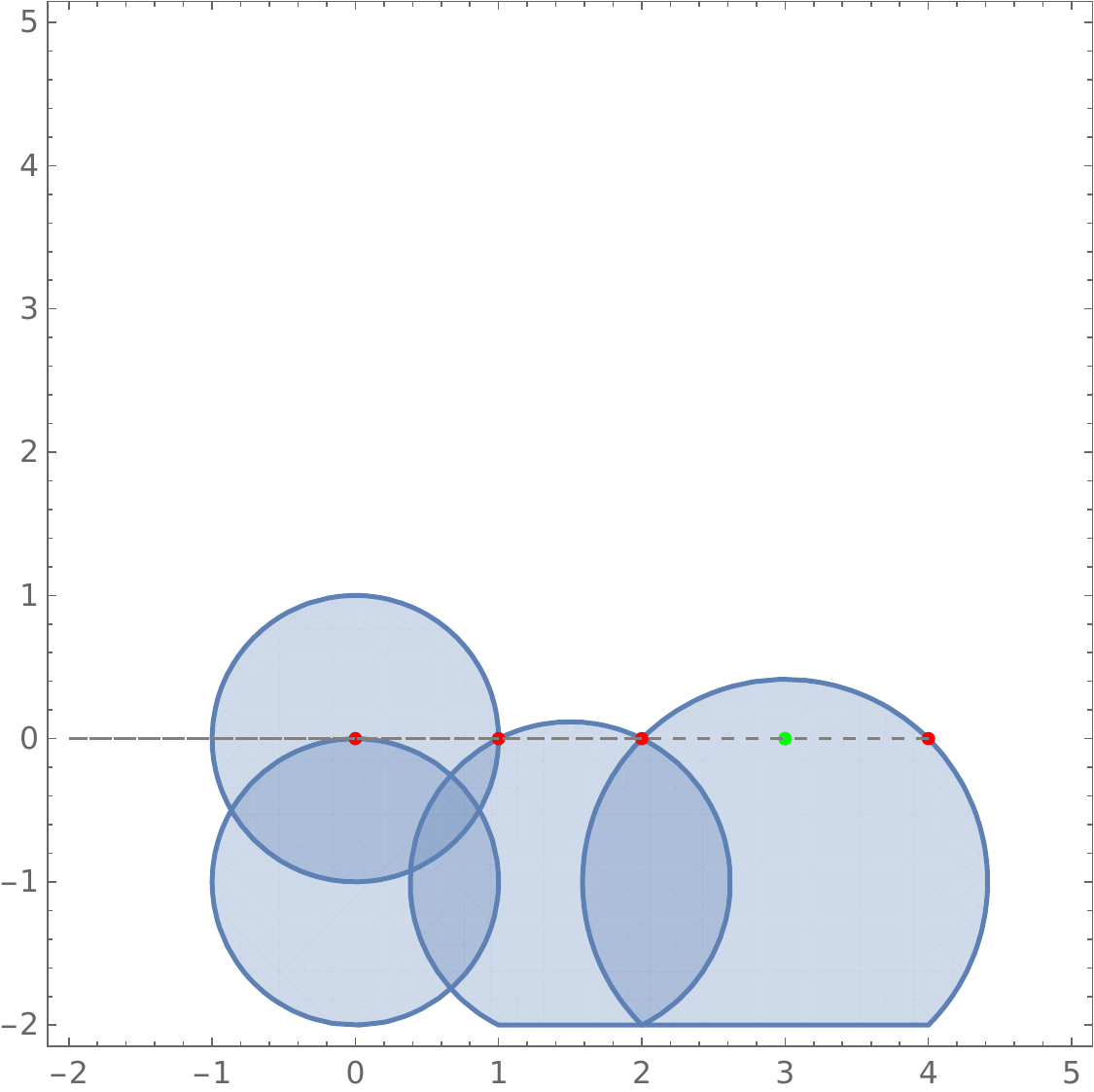}
\end{minipage}\hfill
\begin{minipage}{0.45\textwidth}
\includegraphics[width=\textwidth]{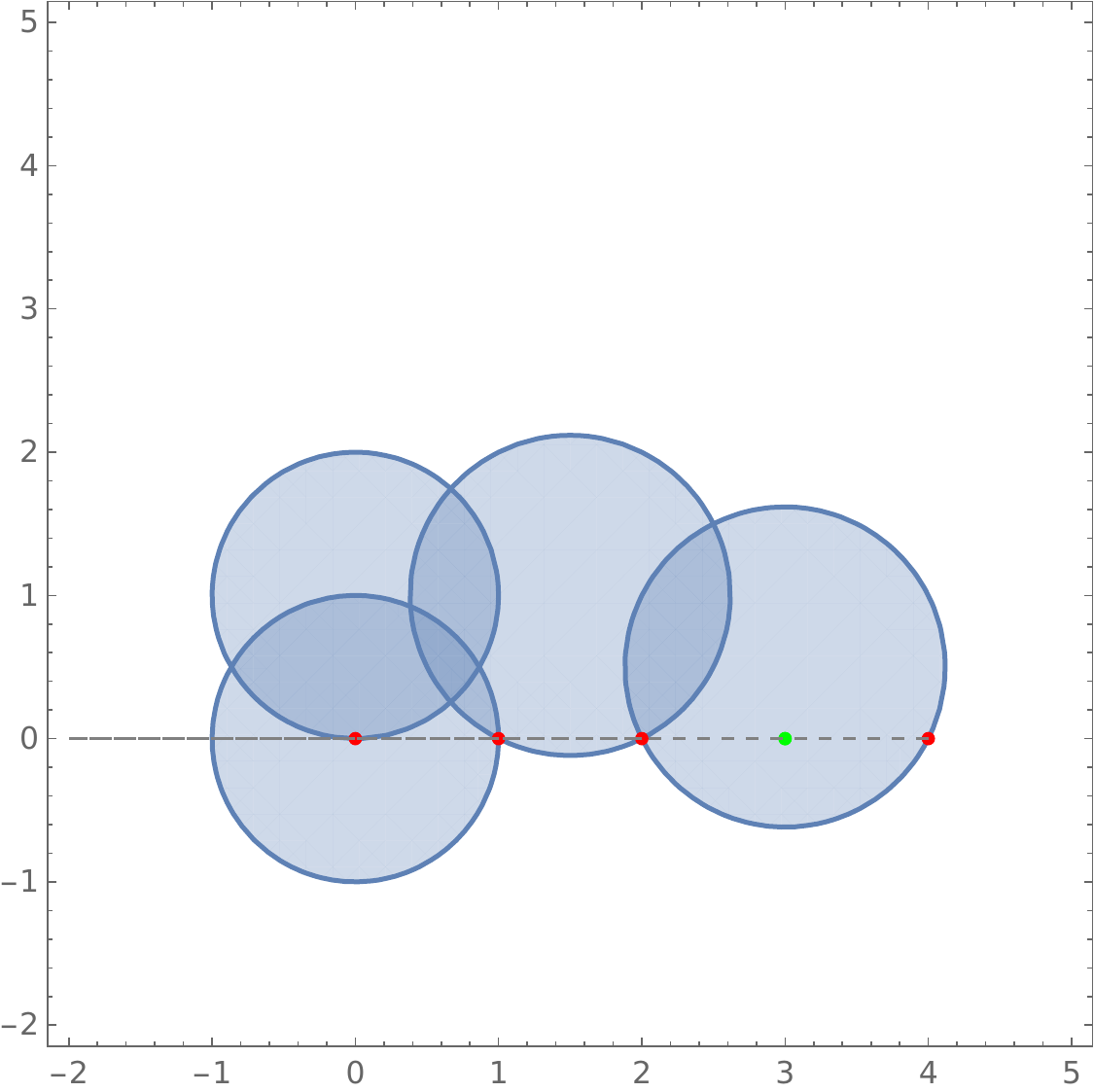}
\end{minipage}
\caption{Analytical continuation with option $\texttt{DeltaPrescription} \rightarrow -I$ on the left and option $\texttt{DeltaPrescription} \rightarrow +I$ on the right. The singular points are shown in red and the end point in green, the staring point is at the origin and shaded circles represent convergence regions. Horizontal dotted lines indicate branch cuts of singular points.}
\label{fig:ContrDirection}
\end{figure}

The \texttt{NExpandHypergeometry} function has the following set of additional options:
\begin{itemize}  
   \item \texttt{SimpleAnalyticContinuation} - can take the values \texttt{True} or \texttt{False}. The default value is \texttt{False}. If \texttt{False}, uses the more complex four-step analytic continuation scheme described in \cite{npb}. If \texttt{True}, uses a simpler two-step scheme in which we only require that $\kappa_i$ in Eq.~\eqref{eq:kappas} be real numbers, rather than real non-negative numbers as in the four-step scheme. This scheme may be more useful in some cases and for comparing results with an answer from polylogarithms if the latter is possible.
   \item \texttt{UseParallelComputing} - can take the values \texttt{True} or \texttt{False}. The default value is \texttt{True}. If \texttt{True}, uses all available kernels for calculations. Parallelization is implemented using the built-in function \texttt{ParallelTable}. Otherwise, it will simply execute sequentially.
   \item \texttt{FrobeniusNumberTerms} - can take the values \texttt{"Auto"} or non-negative integer.       The default value is \texttt{"Auto"}.  
   Specifies the number of elements of the Frobenius series that the program needs to calculate. In the case of \texttt{"Auto"}, the program determines this number automatically based on the desired accuracy.
   \item \texttt{InternalPrecision} - can take the values \texttt{"Auto"} or non-negative integer.       The default value is \texttt{"Auto"}.  
   Specifies the precision with which the individual coefficients in the Frobenius expansion must be calculated. This number must be significantly greater than the required precision in order to obtain a reliable final answer. In the case of \texttt{"Auto"}, the program determines this number automatically based on the desired accuracy.
   \item \texttt{DeltaPrescription} - can take the values $\pm I$. The default value is $-I$. Determines on which side of the cut the end point is taken if its value pertains to the cut. The analytical continuation procedure in these two cases is different. The corresponding example is shown in Figure \ref{fig:ContrDirection}.  
   
\end{itemize} 

\begin{figure}[h!]
\centering
\begin{minipage}{0.45\textwidth}
\includegraphics[width=\textwidth]{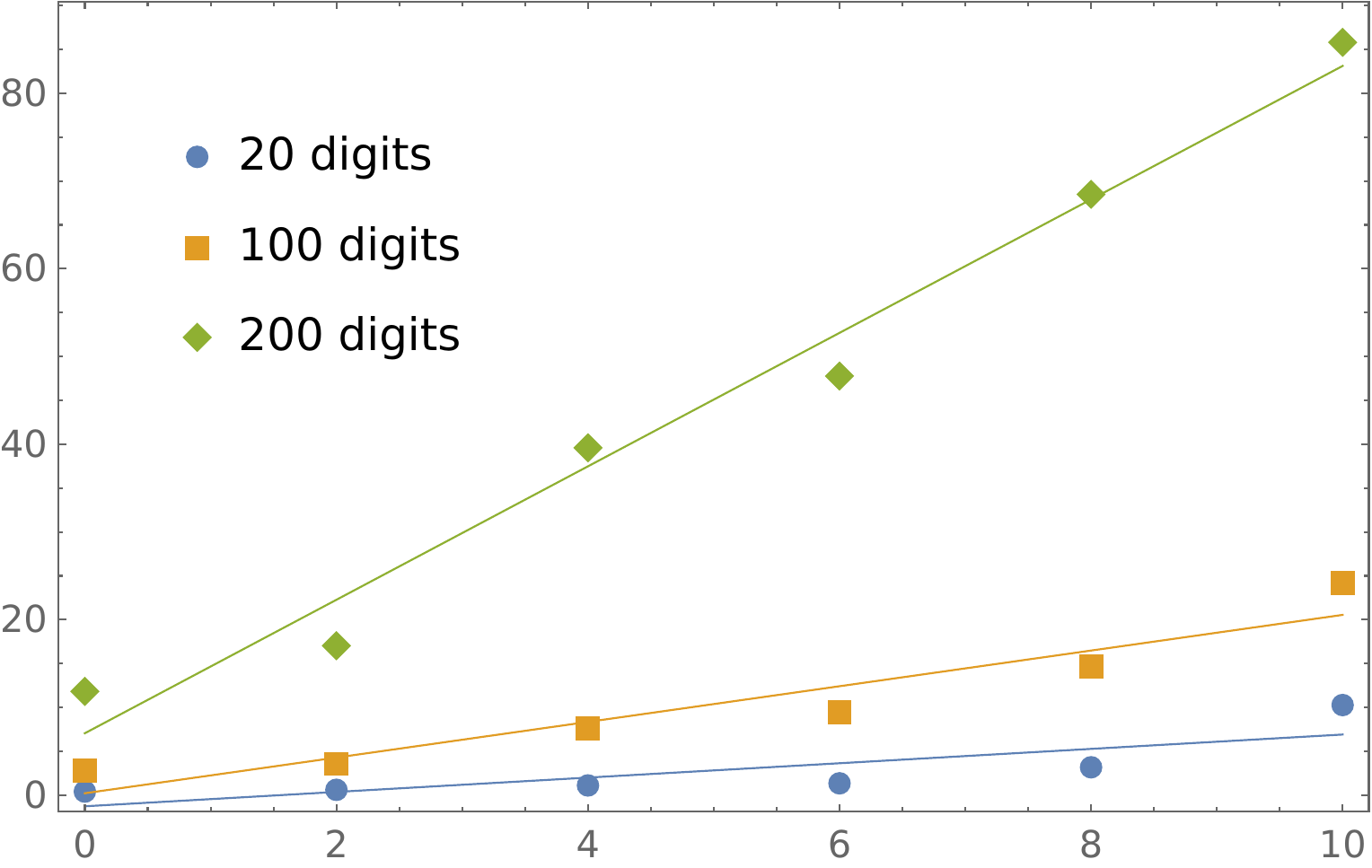}
\end{minipage}\hfill
\begin{minipage}{0.45\textwidth}
\includegraphics[width=\textwidth]{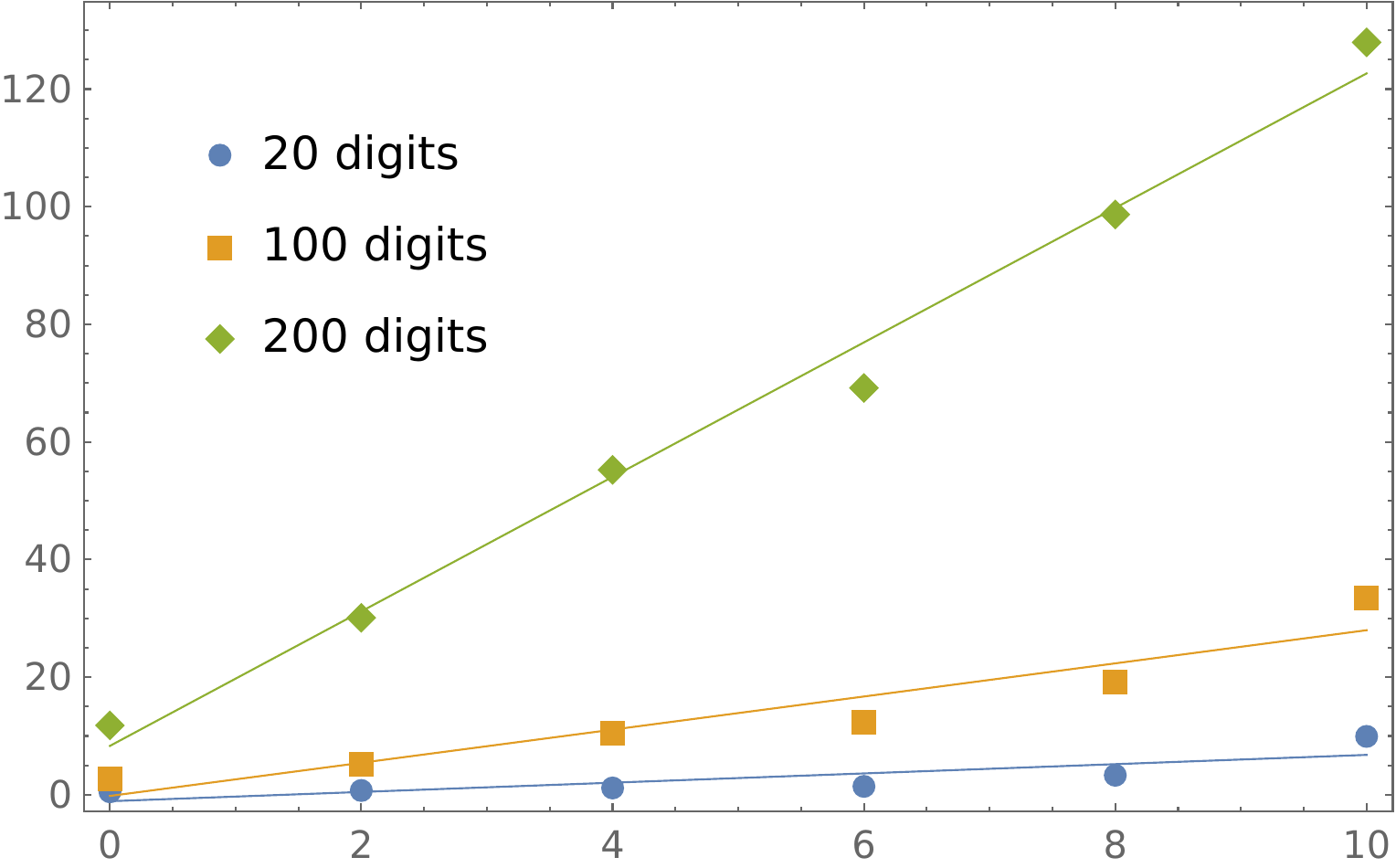}
\end{minipage}
\caption{Average time in seconds required to expand  $F_1\left(\frac{1}{2};1,\ep;\frac{3}{2};\frac{4}{3},\frac{7}{4}\right)$ using 16 parallel kernels on the left and 8 parallel kernels on the right. The horizontal axis shows the number of terms in $\ep$, and the vertical axis shows the evaluation time in seconds. Solid lines represent the best linear fit.}
\label{fig:test1}
\end{figure}

\begin{figure}[h!]
\centering
\begin{minipage}{0.45\textwidth}
\includegraphics[width=\textwidth]{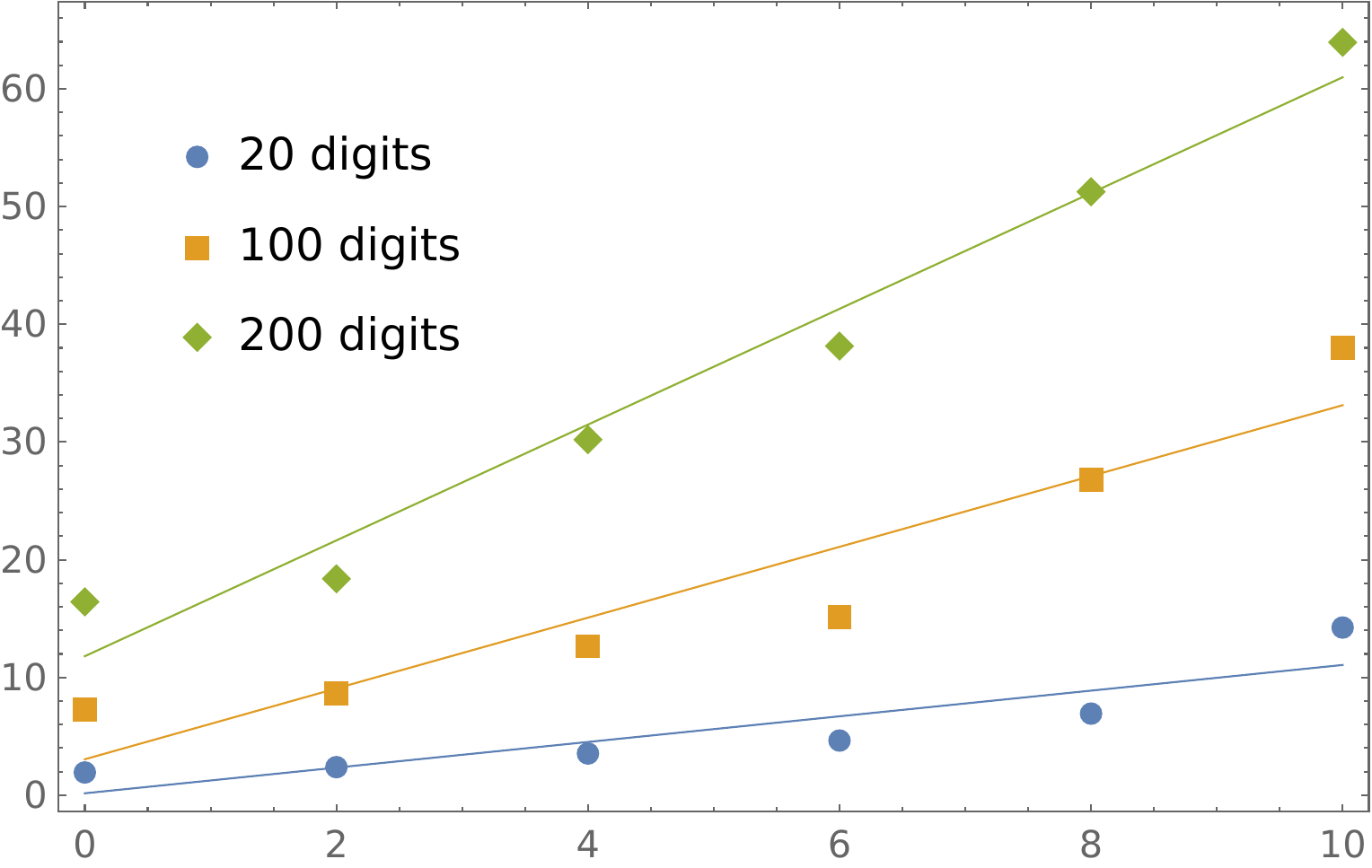}
\end{minipage}\hfill
\begin{minipage}{0.45\textwidth}
\includegraphics[width=\textwidth]{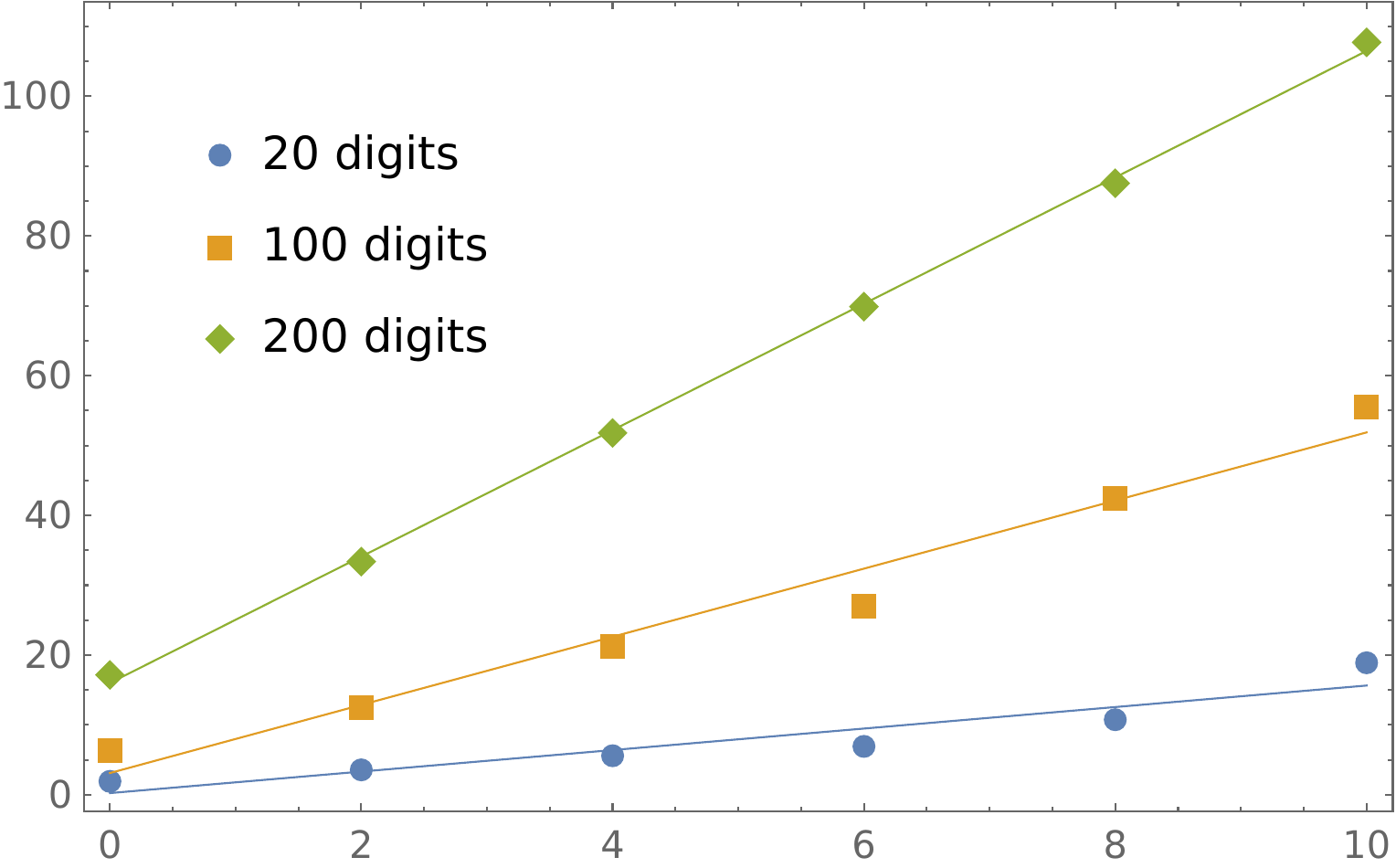}
\end{minipage}
\caption{Average time in seconds required to expand  $F_2\left(1,\frac{2}{3}\ep,1,1+\frac{3}{2}\ep,1-\frac{15}{7}\ep;\frac{3}{2},4\right)$  on the left and $F_D^{(3)}(\frac{1}{2}-\ep; 1,\ep,\ep; 1+2\ep; \frac{4}{3},\frac{3}{4},\frac{8}{5})$ on the right using 8 parallel kernels. The horizontal axis shows the number of terms in $\ep$, and the vertical axis shows the evaluation time in seconds. Solid lines represent the best linear fit.}
\label{fig:test2}
\end{figure}

The computation time for certain functions is also of particular interest. Examples of these test calculations are provided in Figs.~\ref{fig:test1} and \ref{fig:test2}. The test calculations were performed on a standard laptop with a 12th Gen Intel® Core™ i7-12700H × 20 processor. 
The most obvious consequence is that, as the number of terms in the $\ep$ expansion increases, the program’s execution time grows linearly. This is a direct result of our approach to $\ep$ dependencies.

\section{Conclusion}
\label{sec:Conclusion}

In this work, we presented \texttt{PrecisionLauricella}, a software package written in Wolfram Mathematica, for the high-precision numerical evaluation of Lauricella functions with indices linearly dependent on a parameter $\ep$. The package automates the computation of Laurent series expansions of these functions about $\ep = 0$, leveraging their analytical continuation in terms of Frobenius generalized power series. Unlike multi-dimensional series arising in other methods, these one-dimensional series offer significant advantages in terms of computational efficiency and precision, making them especially suitable for large-scale calculations.

Our approach includes a dedicated treatment of the $\ep$ dependences in these expansions, which not only accelerates computations but also facilitates parallelization, making the package scalable for high-performance applications. We validated the accuracy and efficiency of \texttt{PrecisionLauricella} by comparing its results, where applicable, with existing tools for hypergeometric functions, such as those presented in Refs.~\cite{Ananthanarayan:2021bqz,Bera:2024hlq} and Refs.~\cite{Bera:2023pyz,Bezuglov:2023owj}. These comparisons demonstrated the robustness and reliability of the presented approach.

We hope that \texttt{PrecisionLauricella} will serve as a reliable and efficient tool for researchers working with Lauricella and similar hypergeometric functions. We also look forward to exploring further extensions of the package to other classes of hypergeometric functions and additional applications in both mathematics and physics.

\section*{Acknowledgments}
We would like to thank V.V.~Bytev, R.N.~Lee and A.V.~Kotikov  for interesting and stimulating discussions. The work of A.I.O. was supported by the Russian Science Foundation through Grant No.\ 20-12-00205.
The work of M.A.B. and B.A.K. was supported by the German Research Foundation DFG through Grant No.\ KN 365/16-1.
The work of O.L.V. was supported by DFG Research Unit FOR 2926 through Grant No.\ KN 365/13-2.

\bibliographystyle{elsarticle-num}
\bibliography{litr}







\end{document}